\documentclass[useAMS,usenatbib]{mn2e}
\usepackage{epsf}

\newcommand{\bm}[1]{\mbox{\boldmath$#1$}}
\newcommand{\cvir}{c_{\rm vir}}
\newcommand{\rvir}{r_{\rm vir}}
\newcommand{\pdf}{{\rm PDF}}

\title[Statistical distribution of gravitational-lensing excursion angles]
{Statistical distribution of gravitational-lensing excursion angles: 
Winding ways to us from the deep universe}
\author[T. Hamana et al.]
{Takashi Hamana$^{1}$, Matthias Bartelmann$^2$, Naoki Yoshida$^{3}$, 
Christoph Pfrommer$^4$\\
$^1$ National Astronomical Observatory of Japan, 
Mitaka, Tokyo 181-8588, Japan\\
$^2$ ITA, Universit\"at Heidelberg, Tiergartenstr.~15, D--69121 
Heidelberg, Germany\\
$^3$ Department of Physics and Astrophysics, Nagoya University,
Nagoya 464-8602, Japan \\
$^4$ Max-Planck-Institut f\"ur Astrophysik,
Karl-Schwarzschild-Strasse 1, D--85748 Garching, Germany}
\date{Accepted ******; Received ******; in original form 2004 July 29}
\pagerange{\pageref{firstpage}--\pageref{lastpage}} 
\pubyear{2004} 
\volume{000}

\begin{document}

\label{firstpage}

\maketitle

%%%%%%%%%%%%%%%% Abstract
\begin{abstract}
We investigate statistical distributions of differences in 
gravitational-lensing deflections between two light rays, 
the so-called lensing excursion angles.
A probability distribution function of the lensing 
excursion angles, which plays a key role in estimates of lensing effects
on angular clustering of objects (such as galaxies, QSOs and also
the cosmic microwave background temperature map), is known to consist of 
two components; a Gaussian core and an exponential tail.
We use numerical gravitational-lensing experiments in a 
$\Lambda$CDM cosmology for quantifying these two components.
We especially focus on the physical processes
responsible for generating those two components.
We develop a simple empirical model for the exponential tail
which allows us to explore its origin.
We find that the tail is generated by the coherent lensing scatter by 
massive halos with $M>10^{14}h^{-1}M_\odot$ at $z<1$ and that its
exponential shape arises due to the exponential cut-off 
of the halo mass function at that mass range.
On scales larger than 1~arc minute, the tail does not have a practical
influence on the lensing effects on the angular clustering.
Our model predicts that the coherent scatter may have non-negligible
effects on angular clustering at sub-arcminute scales.
\end{abstract}

%%%%% keywords

%%%%%%%%%%%%%%% sec 1 Introduction
\section{Introduction}
Light rays are deflected when they propagate through an inhomogeneous 
gravitational field, such as the real universe we live in.
The lensing deflection angle varies from one direction to another,
and thus the difference in deflection angles between 
two light rays, which we call the ``lensing excursion angle'',
does as well.
Consequently, it is not easy to infer the transverse distance 
between two celestial objects at a cosmological distance from 
their angular separation in the sky.
Strictly speaking, lacking complete knowledge of the matter distribution 
in the universe, this is impossible to do.

Since distance is one of most fundamental physical quantities,
the lack of a precise distance measure to the far universe may
prevent us from a detailed understanding of the universe.
A well known case is that angular correlations of distant galaxies 
and of the temperature map of the cosmic microwave background 
are altered by lensing deflections (see \S 9 of Bartelmann \& 
Schneider 2001 for a review and references therein).

Although we cannot know the lensing excursion angle for an individual
pair of light 
rays, knowledge of their statistical distribution greatly helps
us in estimating the order of magnitude of lensing effects.
In addition, it allows the intrinsic 
angular correlation functions to be deconvolved from measured
correlation functions
(Bartelmann \& Schneider 2001).
It is thus of fundamental importance to understand in detail 
the statistical distribution of the lensing excursion angles.

The analytic model for computing the variance of lensing excursion angles 
in a framework of modern cosmological models was developed by Seljak 
(1994; 1996), based on the linear perturbation theory
(the so-called power spectrum approach).
Hamana \& Mellier (2001) performed numerical experiments of the gravitational
lensing deflections in cold dark matter models and examined the statistical
properties of the lensing excursion angles.
They found that the probability distribution function (PDF) of the 
excursion angles consists of two components, a Gaussian core and 
an exponential tail, and that the variance of the Gaussian core component
agrees well with the prediction by the power spectrum approach.
They argued that the exponential tail may be generated by coherent lensing 
scattering by massive halos which is not taken into account in the power 
spectrum approach.

The purpose of this paper is twofold:
The first is to explore the origin of the exponential tail of the 
lensing excursion angle PDF. 
The second is to develop an empirical model for the exponential tail.
To pursue these purposes, we first examine in detail properties of the
exponential tail
of the excursion-angle PDF using numerical experiments in \S 2.
Then in \S3, we describe a model for the exponential tail which is 
based on the assumption that the tail originates from coherent 
lensing scattering by individual massive dark-matter halos, and compare the 
model predictions with numerical results.
We also discuss a general picture of the light propagation in the 
universe paying special attention to the role of secular random 
deflections by either large or small-scale structures 
and a coherent scatter by a massive halo.
Finally, we give a summary and discussion in \S 4.

%%%%%%%%%%%%%%% sec 2 Ray-tracing simulation
\section{Ray-tracing simulation}

%%%%%%%%%%%%%%% sec 2.1 VLS $N$-body simulation
\subsection{VLS $N$-body simulation}
We performed weak lensing ray-tracing experiments in a Very Large 
$N$-body Simulation (VLS) carried out by the Virgo Consortium 
(Jenkins et al.~2001, and see also Yoshida, Sheth \& Diaferio 2001 
for simulation details).
The simulation was carried out using a parallel P$^3$M code 
(MacFarland et al.~1998) with a force softening length of 
$l_\mathrm{soft}\sim 30\,h^{-1}\mathrm{kpc}$.
The simulation employed $512^3$ CDM particles in a cubic box of 
$479\,h^{-1}\mathrm{Mpc}$ side length, which gives a particle mass 
of $m_\mathrm{part}=6.86\times 10^{10}h^{-1}M_\odot$.
It uses a flat cosmological model with a matter density 
$\Omega_0=0.3$, a cosmological constant $\Omega_\Lambda=0.7$, and a 
Hubble constant $H_0=100h$ with $h=0.7$. 
The initial matter power spectrum was computed using CMBFAST 
(Seljak \& Zaldarriaga 1996) assuming a baryonic matter density of
$\Omega_\mathrm{b}=0.04$. 
The normalization of the power spectrum is taken as $\sigma_8=0.9$.

%%%%%%%%%%% sec 2-2 Weak lensing ray-tracing simulation
\subsection{Weak lensing ray-tracing simulation}

The multiple-lens plane ray-tracing algorithm we used is detailed in
Hamana \& Mellier (2001; see also Bartelmann \& Schneider 1992 and
Jain, Seljak \& White 2000; Hamana, Martel, Futamase 2000; 
Vale \& White 2003 and Hamana, Takada \& Yoshida 2004 for the 
theoretical basics and technical issues); thus in the 
following we describe only aspects specific to the VLS simulation data 
and to ray-tracing experiments in this study.

We use thirteen snapshot outputs from two runs of the $N$-body simulation 
which differ only in the realization of the initial fluctuation field.
A stack of these outputs provides the density field from $z=0$ to $z=6.8$.
We do not use further higher redshift outputs because of two reasons;
(1) discreteness effects of particles (Hamana, Yoshida \& Suto 2002), 
and (2) an artificial power excess in the density power spectrum due to the 
``glass'' initial condition (White 1996) at around the mean separation 
length of particles, both of them are significant at such high redshifts.
For higher redshifts up to the last scattering surface ($z \approx 1100$), 
we simply consider a homogeneous density field.
Thus within $6.8<z<1100$ rays propagate as in a
perfectly homogeneous universe.
This treatment misses lensing contributions from structures at that redshift 
range.
It has turned out that this approximation causes only a minor effect.
We will discuss its influences on our analyses later.

Each $N$-body box is divided into 4 sub-boxes with an equal thickness of 
$119.75\,h^{-1}\mathrm{Mpc}$.
The $N$-body particles in each sub-box are projected onto
{\it lens planes}. In this way, the particle distribution between an 
observer and $z = 6.8$ is projected onto $50$ lens planes. 
Note that, in order to minimize the
difference in redshift between a lens plane and an output of $N$-body
data, only one half of the data (i.e.~two sub-boxes) of $z=0$ output is
used.
The particle distribution on each plane is converted into the surface
density field on a $2048^2$ regular grid using the triangular shaped 
cloud (TSC) assignment scheme (Hockney \& Eastwood 1988). 
The grid size is 0.23$h^{-1}$Mpc which is chosen to maintain the resolution 
provided by the $N$-body simulation and removing at the same time the 
shot noise due to discreteness in the $N$-body simulation (this choice is 
equivalent to the ``large-scale smoothing'' in M\'enard et al., 2003, we 
refer the reader to this reference for further examination of the effective 
resolution of the ray-tracing simulation). 
Its computation follows the procedure described in Hamana \& Mellier (2001).

Having produced surface density fields on all lens planes, $1024^2$
rays are traced backwards from the observer's point using the
multiple-lens plane algorithm (e.g.~Schneider, Ehlers \& Falco
1992). The initial ray directions are set on $1024^2$ grids with a
grid size of $0.25\,\mathrm{arcmin}$, thus the total area covered by
rays is $4.27^2\,$square degrees.
We produced 36 realizations of the underlying density field by 
randomly shifting the simulation boxes in
the direction perpendicular to the line-of-sight using the periodic
boundary conditions of the $N$-body boxes.

%%%% Figure 1
\begin{figure}
\begin{center}
\begin{minipage}{8.2cm}
\epsfxsize=8.2cm 
\epsffile{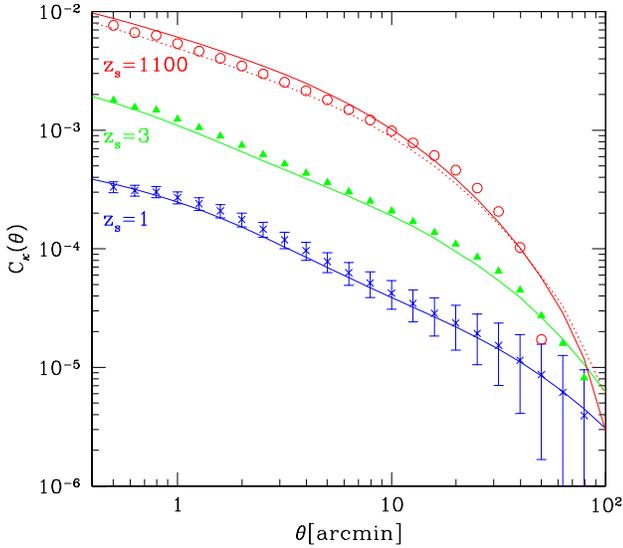}
\end{minipage}
\end{center}
\caption{The convergence two-point correlation functions.
Measurements from ray-tracing experiments are shown by symbols with 
error bars which represent the mean and root-mean-square among the 36 
realizations.
Crosses, filled triangles and open circle are for $z_s=1$, 3 and 1100, 
respectively.
The error bars of the cases for $z_s=3$ and 1100 are not displayed for 
clarity but are as similar magnitude as the $z_s=1$ case.
The solid lines show the theoretical prediction (e.g., Jain \& Seljak 1997)
in which the fitting function of nonlinear power spectrum by Peacock \& 
Dodds (1996) is used to include the effect of the nonlinear growth of the 
density field.
The dotted line shows the theoretical prediction for $z_s=1100$ but
the contribution from density fluctuation at $6.8<z<1100$ is not integrated.}
\label{fig:tpcf}
\end{figure}

The 36 realizations are not perfectly independent because
they are generated from the same $N$-body outputs (but using different 
combinations of random lines of sight) which come from two runs of $N$-body 
simulation.
Therefore the generated lensing data (the lensing deflection field, lensing 
convergence and shear map) are subject to sample variance.
In order to test its magnitude,
we compare the convergence two-point correlation function with its
theoretical prediction (Jain \& Seljak 1997) in Figure \ref{fig:tpcf}.
The measurements from the ray-tracing experiment are plotted by symbols with 
error bars which represent the mean and root-mean-square among the 36 
realizations, while the solid lines show the prediction.
Note that the measurement for $z_s=1100$ should be compared with the 
dotted line which shows the theoretical prediction for $z_s=1100$ but 
the contribution from the density fluctuations between $z=6.8$ and 1100
is ignored.
The measurements are in good agreement with the theoretical prediction
in shape, but are slightly higher in amplitude.
This excess may be mostly attribute to the sample variance and
implies that there exists an excess power in the lensing 
potential field.
Since lensing deflections result from the same potential field,
it is expected that there exists, to a similar extent, an excess in 
the deflection angle statistics.
On scales smaller than 1 arcmin, the slope of the measured 
correlation function becomes flatter than predicted; this is due to the 
limited resolution of the $N$-body simulation.
The effective angular resolution of the convergence field is about 1 arcmin 
for lower redshift ($z_s<3$) and is slightly better for higher redshifts
(see M\'enard et al., 2003 for further discussion on the resolution issue).

%%%%%%%%%%%%%%% sec 2.3 PDF of the lensing excursion angles
\subsection{PDF of the lensing excursion angles}

Using weak-lensing experiments, 
we study the statistics of differences in 
deflection angles between 
two light rays, which we refer to as the ``lensing excursion angle''.
The deflection angle of a light ray, which is computed by the lens equation,
is simply the difference between its positions $\theta_I$ and 
$\theta_S$ on the image and source planes, respectively.
Denoting the deflection angle of two rays  
by $\bmath{\alpha}^1$ and $\bmath{\alpha}^2$, respectively, 
we write the lensing excursion angle between these rays as
$\delta \bmath{\alpha}= \bmath{\alpha}^1 - \bmath{\alpha}^2$.
Similarly we denote  their intrinsic separation by 
$\theta_{12}=|\bmath{\theta}_S^1 - \bmath{\theta}_S^2|$.

%%%% Figure 2
\begin{figure}
\begin{center}
\begin{minipage}{8.2cm}
\epsfxsize=8.2cm 
\epsffile{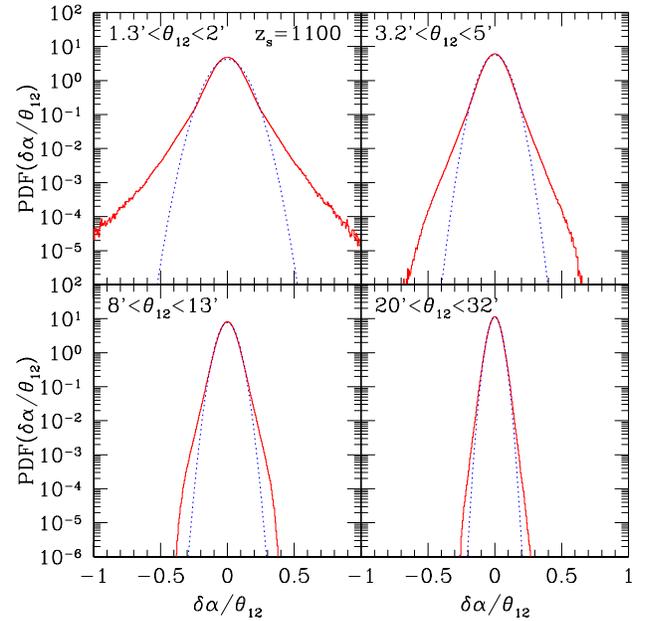}
\end{minipage}
\end{center}
\caption{The solid curves show the probability distribution function of the 
lensing excursion angles normalized by the intrinsic ray separation.
The source redshift is $z_s=1100$.
The ranges of the intrinsic separation of the light-ray pairs
$\theta_{12}$ are
given in each panel.
The dotted curves show Gaussian distributions with their $\sigma$
computed from the measured PDFs (i.e., $\sigma^2=\int dx~x^2 {\rm PDF}(x)$).}
\label{fig:simpdf}
\end{figure}

Let us first look into the probability distribution function (PDF) of 
the lensing 
excursion angles which is one of most fundamental statistics. 
Figure \ref{fig:simpdf} shows the PDFs of the lensing excursion angles 
normalized by its intrinsic separation (i.e.~$\delta \alpha /\theta_{12}$). 
Since the vector field $\bmath{\delta \alpha}$ has no preferred direction, 
we use both components, $\delta \alpha_1$ and $\delta \alpha_2$, 
to compute the PDFs.
The dotted lines in each plot of Figure \ref{fig:simpdf} show the 
Gaussian PDF with its standard dispersion ($\sigma^2$) computed from the 
PDF itself (i.e. $\sigma^2=\int dx~x^2 {\rm PDF}(x)$).
As was first pointed out by Hamana \& Mellier (2001), the PDFs consist of
two components, a Gaussian core and the exponential tail, which are 
generated by different physical processes as we will discuss below.

%%%% Figure 3
\begin{figure}
\begin{center}
\begin{minipage}{8.2cm}
\epsfxsize=8.2cm 
\epsffile{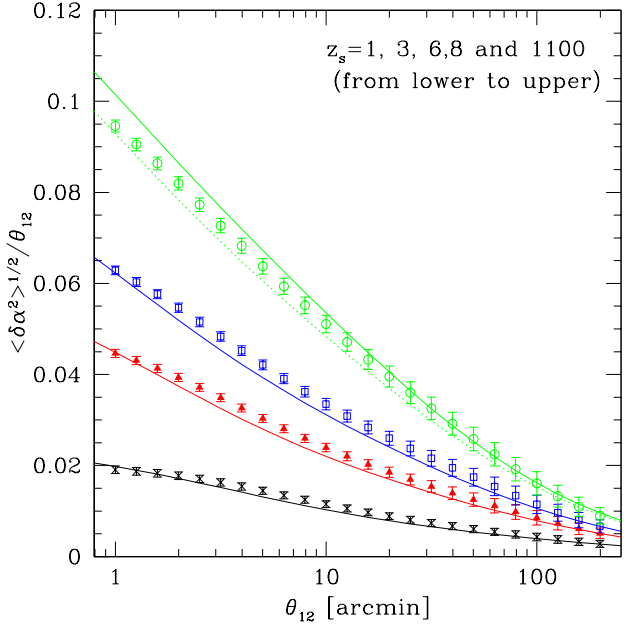}
\end{minipage}
\end{center}
\caption{The standard dispersion of the lensing excursion angles 
plotted as the function of the separation angle.
Symbols and error bars show the mean and RMS among 36 
realizations of the ray-tracing numerical experiments.
Solid lines represent the theoretical predictions from the power spectrum 
approach (Seljak 1996), in which the fitting function of the nonlinear 
power spectrum by Peacock \& Dodds (1996) was used.
The source redshifts are, from lower to higher, $z_s$=1, 3, 6.8 and 1100,
respectively. 
The dotted line shows the theoretical prediction for $z_s=1100$ but
the contribution from density fluctuation at $6.8<z<1100$ is ignored.}
\label{fig:variance}
\end{figure}

The origin of the Gaussian core is explained as follows:
Light rays from a cosmological distance undergo many (either strong or weak)
gravitational lensing deflections.
Since the spatial distribution of lenses at a large separation are 
uncorrelated, rays basically undergo many uncorrelated deflections.
Provided the separation between two rays is so large that
effects of coherent scattering can be ignored, two light rays 
undergo independent deflections.
According to the central limit theorem, the statistical distribution of 
the lensing excursion angles of such light-ray pairs is given by a
Gaussian.
A necessary condition for the central limit theorem to hold is that
the parent distribution of the individual events which are being
superposed has finite variance.
The deflection angle calculated in the weak-lensing regime using the
power-spectrum approach has finite variance (Seljak 1994; 1996), but
it is based on linearized gravity and ignores lensing by individual
halos.
On the other hand, numerical gravitational lensing experiments show
that the PDF consists of the Gaussian core and the exponential
tail. In order to understand whether the Gaussian core can indeed be
caused by the superposition of many deflections, we need to
investigate the variance of the excursion angle.
In particular, the numerical experiments miss the influence of
numerous distant lenses,
because of their necessarily finite volume.
We will now show, using a simple approach, that 
accumulated contributions from very distant lenses do not
significantly
affect the excursion angle variance, thus it remains finite.

Consider first a single light ray passing the lens plane in the
origin. There is a
finite number of lenses close to the ray, thus we can restrict
ourselves to distant lenses since we are investigating whether the
deflection-angle variance is finite or not. Axially symmetric lenses
more distant than their (e.g.~virial) radii act as point lenses, thus
we can approximate their deflection angles by
$\vec\alpha_i=\vec\theta_i/\theta_i^2$. Assuming the lenses have a
number density $n$ and are randomly distributed, the variance of the
total deflection angle contributed by lenses in a ring around the
origin with radius $\theta$ and width $d\theta$ is
\begin{equation}
  \left\langle\vec\alpha^2\right\rangle=
  \left\langle\sum_{i=1}^N\frac{1}{\theta^2}\right\rangle=
  \frac{2\pi n\theta\,d\theta}{\theta^2}=
  2\pi n\,d\ln\theta\;.
\label{eq:var1}
\end{equation}
Integrating over $\theta$ shows that the variance diverges
logarithmically.

The situation changes for the excursion angle. Consider two light rays
piercing the lens plane at positions $\vec\theta_{1,2}=(\mp
d/2,0)$. Specializing again to distant lenses, we can approximate the
individual deflection angles by those of point lenses. The excursion
angle of the lenses in a ring of radius $\theta$ and width $d\theta$
around the origin can then be expanded to lowest order in $d/\theta$,
\begin{equation}
  \delta\vec\alpha(\theta)=
  \sum_{i=1}^N\frac{d}{\theta^2}
  \left(-\cos2\phi_i,\sin2\phi_i\right)\;,
\label{eq:var2}
\end{equation}
where $\phi_i$ is the polar angle of the $i$-th lens. Again assuming
randomly distributed lenses, the variance of the excursion angle
contributed by the lenses in the ring is thus
\begin{equation}
  \left\langle\delta\vec\alpha^2(\theta)\right\rangle=
  \left\langle\sum_{i=1}^N\frac{\pi d^2}{\theta^4}\right\rangle=
  \frac{2\pi^2d^2n\theta\,d\theta}{\theta^4}\;,
\label{eq:var3}
\end{equation}
i.e.~the excursion-angle variance converges like $\theta^{-2}$ when
integrated over $\theta$ to infinity. Thus, we can apply the central
limit theorem to the excursion angle, while we could not for the
deflection angle itself.

Now we test the theoretical model prediction for the variance of the 
excursion angles developed by Seljak (1994; 1996) against our 
numerical results.
Figure \ref{fig:variance} compares the standard dispersion measured form the 
numerical experiments with the theoretical prediction.
The dispersion becomes larger as the light rays travel a longer distance,
because the rays can undergo more deflections.
It is found in the plot that the measurements are slightly larger than
the prediction. However, a similar excess is seen in the 
convergence correlation function (Figure \ref{fig:tpcf}), thus this
is mostly due to the sample variance.
We may therefore conclude that the power spectrum approach provides
a good prediction even for $z_s=1100$, and the non-Gaussian tail has
no strong contribution to the variance.
It is important to notice that coherent scattering by lensing
due to massive halos that generate the exponential tail contribute
only very little to the excess in the measured dispersion 
over the prediction. 

%%%% Figure 4
\begin{figure}
\begin{center}
\begin{minipage}{8.2cm}
\epsfxsize=8.2cm 
\epsffile{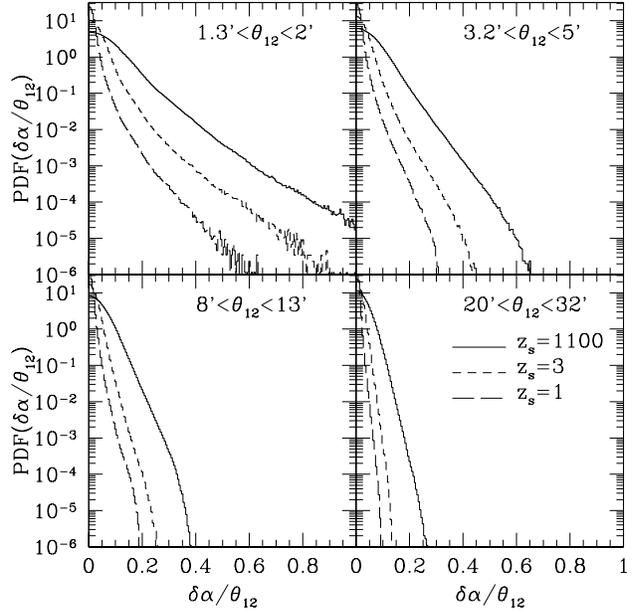}
\end{minipage}
\end{center}
\caption{The PDF of lensing excursion angles normalized by its intrinsic 
separation (plotted positive side only). 
The source redshifts are, from narrower 
to broader PDF, $z_s$=1, 3 and 1100, respectively. 
As these plots show, for a given $\theta$ range, 
the amplitude of the PDF tail becomes higher as $z_s$ increases, while 
their slope is almost unchanged for plotted redshifts $z_s>1$.}
\label{fig:simpdfz}
\end{figure}

%%%% Figure 5
\begin{figure}
\begin{center}
\begin{minipage}{8.2cm}
\epsfxsize=8.2cm 
\epsffile{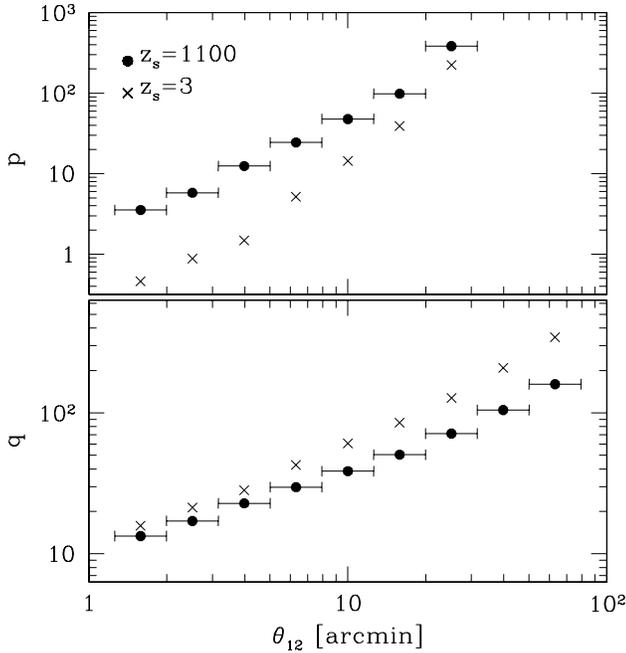}
\end{minipage}
\end{center}
\caption{Parameters in the exponential distribution (eq. \ref{expwing}) 
computed by fitting to the non-Gaussian tail of PDFs from ray-tracing 
experiments.
The filled circles and crosses show the results from the $z_s=1100$ and 
$z_s=3$ cases, respectively.
The error bars represent the ranges of ray separations taken to compute the 
PDFs.}
\label{fig:pqfit}
\end{figure}

Let us now turn to the tail of the lensing excursion angle PDF.
Figure \ref{fig:simpdfz} compares the PDF obtained from the ray-tracing 
numerical experiments for three source redshifts,
$z_s=1$, 3 and 1100, and for various ranges of ray separations.
This Figure represents major characteristics of the non-Gaussian tail:
(a) it has an approximately exponential slope;
(b) it changes little with the source redshift,
but its amplitude increases with the source redshift,
at least within the redshift range we consider ($z_s>1$).
We fit the tail of the PDFs to the exponential distribution:
\begin{equation}
\label{expwing}
E(x)=p\exp(-q x).
\end{equation}
To do this,  we take two points $x_1$ and $x_2$ such that 
$\pdf(x1)=1.0\times10^{-2}$ and $\pdf(x2)=1.0\times10^{-3}$.
The comparison between the results of $z_s=3$ and $z_s=1100$ plotted in 
Figure \ref{fig:pqfit} confirms the 
above point (b) in a quantitative manner.
Note that fitting the exponential function to the PDF tails becomes
poor for large ray separations because the non-Gaussian tail does not 
appear prominently due to the limited statistics.
This accounts for the steep rise of both $p$ and $q$ at large $\theta_{12}$
which rather reflects the slope of the Gaussian core.

%%%% Figure 6
\begin{figure}
\begin{center}
\begin{minipage}{8.2cm}
\epsfxsize=8.2cm 
\epsffile{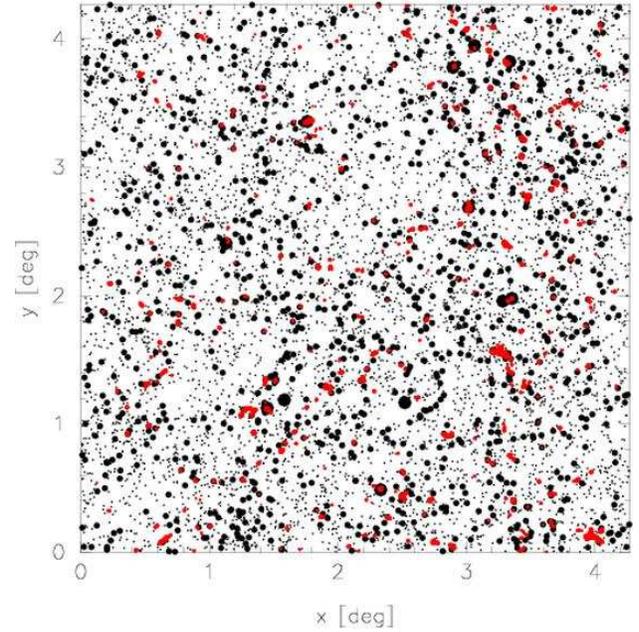}
\end{minipage}
\end{center}
\caption{The celestial distributions of ray pairs having a large lensing
excursion angle and of massive halos in one realization of the numerical
experiment.  See Hamana et al.~(2004) for a detailed description of
construction of the halo catalog on a light-cone.  Black symbols represent
halos; the large filled circles, small filled circles and dots are for halos
with $M_{\rm halo}>4\times 10^{14}$, $4\times 10^{14} > M_{\rm halo}>4\times
10^{13}$ and $4\times 10^{13} > M_{\rm halo}>1\times 10^{13}$
[$h^{-1}M_\odot$], respectively. Only the halos within the redshift interval
between 0 and 1 are displayed.  Red dots represent the ray pairs which obey
the following criteria, an unlensed ray separation of $\theta_{12}>1$ arcmin
and an excursion angle of $|\bm{\delta \alpha}|/\theta_{12} > 0.7$. The
middle points of ray pairs are displayed.}  
\label{fig:rays_halos}
\end{figure}

In the next section, we develop a model of the exponential tail and explore
its origin.  To do this, a visual impression from Figure
\ref{fig:rays_halos} could be informative.  In this figure, the celestial
distributions of ray pairs having a large lensing excursion angle and of
massive halos ($M_{\rm halo} > 10^{13}h^{-1}M_\odot$) in one realization of
the numerical experiment are displayed (see Hamana et al.~2004 for a
detailed description of the construction of a halo catalog on a light-cone).
Only the halos within the redshift interval between 0 and 1 are plotted
using black symbols.  Red dots represent the ray pairs which obey the
following criteria, an unlensed ray separation of $\theta_{12}>1$ arcmin and
an excursion angle of $|\bm{\delta \alpha}|/\theta_{12} > 0.7$.  Apparently,
most of large excursion angle ray pairs pass very close to a massive halo.
We argue in the following section that the exponential tail results from
coherent strong deflections of two nearby rays by a massive halo, and
explain the origin of the above characteristics using simple models.

%%%%%%%%%%%%%%% sec 3 Excursion angle by universal halos
\section{Origin of the exponential tail of the excursion angle PDF}

In this section we explore the origin of the exponential tail 
of the excursion angle PDF found in the ray-tracing numerical experiments.
For this purpose, we focus on the tail part and do not consider the Gaussian 
core whose origin has been investigated in the literature
(Seljak 1994; 1996; see also chapter 9 of Bartelmann \& Schneider) and 
also in the last section.
We develop a theoretical model from two assumptions:
Large excursion angles are mainly caused by the strong lensing of a 
massive halo, and the probability for a ray to undergo multiple 
strong lensing events is negligible.
The former is reasonable because a process that is not taken into
account in the power spectrum approach could generate non-Gaussian 
features.
Also the visual impression from Figure \ref{fig:rays_halos} could be
a support of that idea.
The latter is validated by the observational fact of the small 
cross section for strong lensing events by a single lens (either 
a galaxy or cluster of galaxies) such as multiply-imaged
QSOs and strongly-lensed arc-like images of distant galaxies.
Thus, it is certain that multiple scattering by more than one massive 
halo is very rare.

We consider the same $\Lambda$CDM cosmology as one adopted for the 
numerical experiments in \S 2.
We denote the PDF for finding a ray pair with $\theta_{12}$ having the
excursion angle
${\bm{\delta \alpha}}$ by $\pdf({\bm{\delta \alpha}}|\theta_{12})$.

%%%%%%%%%%%%%%% sec 3.1 Universal density profile halo lens
\subsection{Lensing deflection by a universal density profile halo}
\label{NFWlens}

Navarro Frenk \& White (1996; 1997, NFW hereafter) found from $N$-body 
simulations that the density profile of dark matter halos can be fitted 
by a universal form regardless of their mass and redshift.
We adopt a truncated universal profile;
\begin{equation}
\label{nfw:rho}
\rho(x)={{\rho_s} \over {x^s (1+x)^{3-s}}}, 
\quad x={r\over{r_s}},
\end{equation}
for $r<\rvir$ and 0 otherwise, where $r_s$ and $\rvir$ are the scale 
radius and virial radius, respectively.
It is convenient to introduce the concentration parameter 
$\cvir=\rvir/r_s$.
Navarro et al.~(1996) proposed the universal inner slope of $s=1$, 
while a steeper slope was claimed by later studies using higher resolution 
$N$-body simulations; Moore et al.~(1998; 1999), Ghigna et al.~(2000) and
Fukushige \& Makino (2001; 2003) found larger values
such as $s=1.5$, while Jing (2000) and
Jing \& Suto (2000) pointed out that it varies from 1.1 to 1.5 and argued
a possible weak dependence on the halo mass.
In this paper, we consider two cases $s=1$ and 1.5. 
Navarro et al.~(1997) and Bullock et al. (2001) have extensively 
examined a relation between the concentration parameter and the halo mass
and its redshift evolution adopting a fixed value of $s=1$.
We adopt a generalized mass-concentration relation proposed by 
Oguri, Taruya \& Suto (2001; see also Keeton \& Madau 2001); 
\begin{equation}
\label{cvir}
\cvir(M,z)=(2-s){{c_\ast} \over {1+z}}
\left( {{M}\over {10^{14}h^{-1}M_\odot}}\right)^{-0.13}.
\end{equation}
Bullock et al. (2001) suggested $c_\ast \sim 8$ for the $\Lambda$CDM model,
which we adopt as a fiducial choice.
We note that there is a relatively large scatter in this relation 
(Bullock et al.~2001; Jing 2000).
The virial mass (defined by the mass within the virial radius $\rvir$) 
of the universal halo is given by
\begin{equation}
\label{nfw:Mvir}
M_{\rm vir} = 4\pi \rho_s \rvir^3 {{m(\cvir,s)}\over {\cvir^3}}, 
\end{equation}
with
\begin{equation}
\label{nfw:m}
m(\cvir,s)=\int_0^{\cvir} dx~{{x^{2-s}} \over 
{(1+x)^{3-s}}}.
\end{equation}
Since the spherical collapse model indicates that 
$M_{\rm vir}=4\pi r_{\rm vir}^3\delta_{\rm vir}(z) \bar{\rho}_0/3$, 
where $\delta_{\rm vir}$ is the over-density of collapse
(see Nakamura \& Suto 1997 and Henry
2000 for useful fitting functions), one can express
$\rho_s$ in terms of $\delta_{\rm vir}(z)$, $\cvir$ and $s$:
\begin{equation}
\label{delta_s}
\rho_s={{\delta_{\rm vir} \bar{\rho}_0} \over 3} {{\cvir^3} \over 
{m(\cvir,s)}}.
\end{equation}

\begin{figure}
\begin{center}
\begin{minipage}{8cm}
\epsfxsize=8cm 
\epsffile{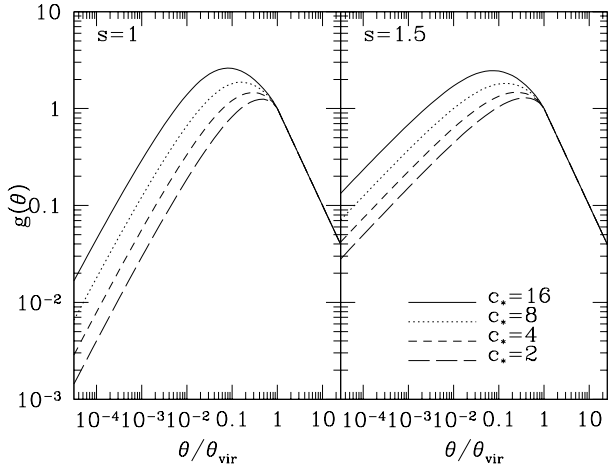}
\end{minipage}
\end{center}
\caption{Deflection angle profiles of the truncated universal density
profile lens, $g(\theta)$ defined in eq.~(\ref{g}), as a function of the 
impact parameter (normalized by the angular virial radius)
for various concentration parameters denoted in the plot.
The left panel is for $s=1$ and right panel for $s=1.5$.
Note that $g(\theta)$ has an asymptotic inner slope of 
$\propto \theta^{0.82}$ ($\propto \theta^{0.48}$) for 
$s=1$ ($s=1.5$).}
\label{fig:g}
\end{figure}

Let us summarize basic equations for gravitational lensing properties 
of the truncated universal profile halo (Takada \& Jain 2003, 
see Bartelmann 1996; Wright \& Brainerd 2000; Oguri et al. 2001 for 
lensing properties of the non-truncated universal profile lens model). 
The surface mass density of the truncated universal profile halo is given by,
\begin{equation}
\label{nfw:Sigma}
\Sigma(y)=\int_{-\sqrt{\cvir^2-y^2}}^{\sqrt{\cvir^2-y^2}} 
dz~\rho({\bm{y}},z) =  
2 \rho_s r_s f(y),\quad y={r\over{r_s}},
\end{equation}
with
\begin{equation}
\label{f}
f(y)=\int_0^{{\sqrt{\cvir^2-y^2}}} dz~{1\over
{(y^2+z^2)^{s/2}(1+\sqrt{y^2+z^2})^{(3-s)}}},
\end{equation}
for $y\le \cvir$ and $f(y)=0$ otherwise.
The projected mass within a radius $b$ is
\begin{eqnarray}
\label{pmass}
M(<b) 
&=& 2 \pi \int_0^b dy'~y' \Sigma(y')\nonumber\\
&=& 4 \pi r_s^3 \rho_s \int_0^h dy~y f(y),
\end{eqnarray}
where $h=b/r_s$. 
We perform the above integration numerically.
The thin lens equation is written by
\begin{equation}
\label{thinlens}
\theta_S=\theta_I-\alpha(D_l \theta_I),
\end{equation}
with
\begin{equation}
\label{alpha}
\alpha(b)={{4GM(<b)} \over {c^2 b}} {{D_{ls}}\over {D_s}}. 
\end{equation}
In the last expression, the origin of the coordinates is taken at 
the lens center, $b=D_l \theta_I$ is the impact parameter, and 
$D_l$, $D_{ls}$ and $D_s$ are the angular diameter distances from 
observer to lens, from lens to source, and from observer to source, 
respectively.
The deflection angle of the truncated universal profile halo is given by,
\begin{equation}
\label{alphaNFW}
\alpha(\theta)=\alpha_\ast g(\theta), 
\end{equation}
with
\begin{equation}
\label{g}
g(\theta)={{\cvir} \over {m(\cvir,s)}}
{{\int_0^x dy~y f(y)} \over x},\quad x={{D_l \theta}\over{r_s}}
\end{equation}
and 
\begin{eqnarray}
\label{alphaast}
\alpha_\ast &=& 2 \Omega_{\rm m} {{D_{ls}}\over {D_s}} 
\left( {{H_0}\over {c}}\right)^2 
\rvir^2  \delta_{\rm vir} \nonumber\\
&\simeq& 4'' \left( {{M_{\rm vir}} \over {10^{14}h^{-1}M_\odot}} \right)
\left( {\rvir \over {1h^{-1}{\rm Mpc}}} \right)^{-1}
{{D_{ls}} \over {D_s}}.
\end{eqnarray}
Note that $\alpha_\ast \propto M^{2/3}$.
It is important to notice that a dependence of the halo profile parameters 
on the deflection angle enters only through the function $g(\theta)$.
Note that for $\theta\ge \theta_{\rm vir}$ it reduces to 
$g(\theta)=\theta_{\rm vir}/\theta$ (where $\theta_{\rm vir}$ is the angular 
virial radius defined by $\theta_{\rm vir}=\rvir/D_l$).
In Figure \ref{fig:g}, the function $g(\theta)$ is plotted for 
various value of $c_\ast$.
As one may see in the Figure, the deflection angle profile $g(\theta)$ 
peaks at $\theta \sim \theta_{\rm vir}/c_\ast = \theta_s$ 
($\theta_s=r_s/D_l$) and 
the peak value does not strongly depend on the inner slope $s$.
It is also found that the peak value relates to
the concentration parameter roughly by $g_{\rm max}\sim 0.1 c_*+1$.
Therefore, in a reasonable range of $c_*$ the maximum deflection angle 
by a single universal halo lens is $\alpha_{\rm max}=(1-3) \alpha_\ast$.
One may also find that at the inner part, the deflection angle is larger 
for a larger $c_\ast$ or for a steeper inner slope 
(thus for more centrally concentrated halos).
We find that $g(\theta)$ has an asymptotic inner slope of 
$\propto \theta^{0.82}$ 
($\propto \theta^{0.48}$) for $s=1$ ($s=1.5$).

%%%% Figure 7

%%%%%%%%%%%%%%% sec 3.2 Lensing excursion angles
\subsection{Lensing excursion angles}
\label{sec:excursion}

Since the deflection angle of a universal density profile halo is finite,
the excursion angle is finite as well.
Clearly, the largest excursion angle is $2\alpha_{\rm max}$ which happens 
when one ray passes at the distance $\sim \theta_s$ from the lens 
and the other ray passes at the same distance in the opposite side of 
the lens.
Thus this happens only if $\theta_{12}=2 \theta_s$, and is very rare.
Let us consider the maximum excursion angle produced for other ray 
separations.
If $\theta_{12} > \theta_s$, the largest excursion angle is in the 
range $\alpha_{\rm max}<\delta\alpha<2~\alpha_{\rm max}$.
This happens when one ray passes at $\sim\theta_s$ and the another ray 
passes at the opposite side of the lens in the direction connecting the 
lens center and the first ray.
While if $\theta_{12} < \theta_s$, the largest excursion angle is smaller
than $\alpha_{\rm max}$.
An important consequence of this is that for ray pairs with the separation 
angle larger than $\theta_s$, the maximum excursion angle does not strongly 
depend on the ray separation but lies in a small range of 
$(1-2)\alpha_{\rm max}$,
and it scales with the halo mass as $\propto M^{2/3}$.

%%%%%%%%%%%%%%% sec 3.3 {PDF of $\delta \alpha$
\subsection{PDF of $\delta \alpha$}
\label{sec:pdf}

Let us first consider the probability distribution induced by one halo,
which we denote by
$\pdf_1({\bm{\delta \alpha}}|\theta_{12})$.
Let $p(\bm{\theta})$ be the probability of a ray passing at a small area
$\bm{\theta}\rightarrow\bm{\theta}+\bm{\delta \theta}$ from a lens center, 
which is given by the cross section area
(denoted by $A$) normalized by the unit solid angle ($d\Omega$):
\begin{equation}
\label{sigmat}
p(\bm{\theta})={
{A(\bm{\theta} \rightarrow \bm{\theta}+\bm{\delta \theta})} 
\over {d\Omega}}.
\end{equation}
Then $\pdf_1({\bm{\delta \alpha}}|\theta_{12})$ is given by the joint
probability,
\begin{equation}
\label{pdf1}
\pdf_1({\bm{\delta \alpha}}|\theta_{12})=
\int {{d^2\theta_1} \over {d\Omega}} 
\int {{d\phi_{12}} \over {2 \pi}} 
p(\bm{\theta}_1)~p(\bm{\theta}_1+\bm{\theta}_{12}),
\end{equation}
where 
$\bm{\theta}_{12}
=\{\theta_{12} \cos(\phi_{12}),\theta_{12} \sin(\phi_{12})\}$.
Note that since we are considering lensing by a single halo having 
a certain density profile and mass, 
given a configuration of a light ray pair, 
its excursion angle is {\it uniquely determined}.
The total $\pdf({\bm{\delta \alpha}}|\theta_{12})$ is obtained 
by summing $\pdf_1(\delta \alpha | \theta_{12})$ over 
halos within a light-cone volume,
\begin{eqnarray}
\label{pdf}
\pdf(\bm{\delta \alpha}|\theta_{12})&=&
\int dV \int dM~n_{\rm halo}(M,z) 
\pdf_1(\delta \alpha | \theta_{12})\nonumber\\
&=& \int_0^{r(z_s)} dr~r^2 \int dM \nonumber\\
&& \times n_{\rm halo}(M,r[z]) 
\pdf_1(\bm{\delta \alpha} | \theta_{12}),
\end{eqnarray}
where $r$ is the comoving radial distance, $dV=r^2 dr$ 
(this expression is valid only for a flat cosmological model) 
is the unit volume element and $n_{\rm halo}(M,z)$
is the halo mass function.
We adopted the mass function by Sheth \& Tormen (1999).
Note that this approach breaks down for 
small excursion angles where secular scattering by distant and/or
small halos are important, which generate the
Gaussian core of the excursion-angle distribution.

%%%%%%%%%%%%%%% sec 3.4 Results
\subsection{Results}

%%%% Figure 8
\begin{figure}
\begin{center}
\begin{minipage}{8.2cm}
\epsfxsize=8.2cm 
\epsffile{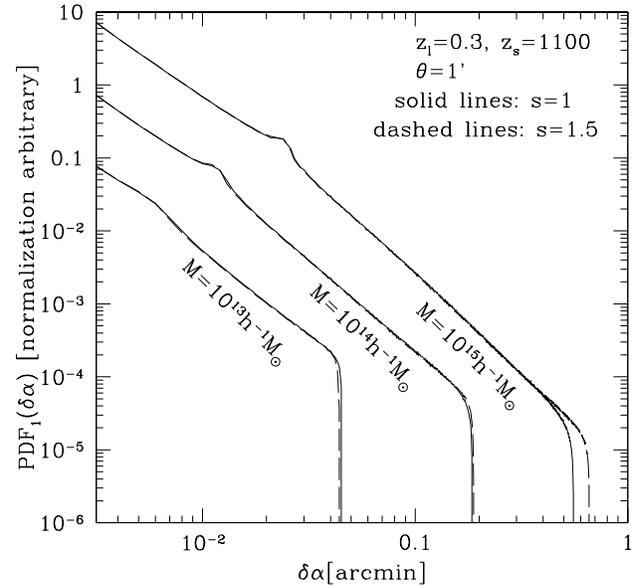}
\end{minipage}
\end{center}
\caption{PDF of the lensing excursion angles from one halo, 
$\pdf_1(\delta \alpha)$, for three halos masses
$M_{\rm halo}=10^{13}$, $10^{14}$ and $10^{15}h^{-1}M_\odot$.
The normalization is arbitrary.
The lens and source redshifts are $z_l=0.3$ and $z_s=1100$, respectively.
The solid lines are for $s=1$ and dashed lines are for $s=1.5$.
The concentration parameter taken is $c_*=8$ for all cases. }
\label{fig:pdf1}
\end{figure}

Let us start with the excursion angle PDFs from one halo
plotted in Figure \ref{fig:pdf1} which help to understand the origin of 
the exponential tail.
The most important point which should be noticed is the sharp cut-off 
in a large excursion angle.
This is a natural consequence of the fact that the deflection angle 
of the universal density profile halo is finite (see \S \ref{NFWlens}).
The maximum excursion angle scales with the halo mass roughly as 
$\propto M^{2/3}$ (with a small correction by the mass dependence 
of the concentration parameter) as far as the separation angle is larger 
than the angular scale radius ($\theta_s$) of a lensing halo, 
as explained in the last subsection.
The other important point is that the mass-independent double power-law 
slope, its power-law slope is $\propto \delta\alpha^{-2}$ for smaller 
excursion angles and $\propto \delta\alpha^{-2.64}$ for larger angles. 
The former is generated by ray pairs in which both rays pass outside of the 
virial radius, while the latter is generated by pairs of rays of which
one passes outside of the halo, and the other inside.

%%%% Figure 9
\begin{figure}
\begin{center}
\begin{minipage}{8.2cm}
\epsfxsize=8.2cm 
\epsffile{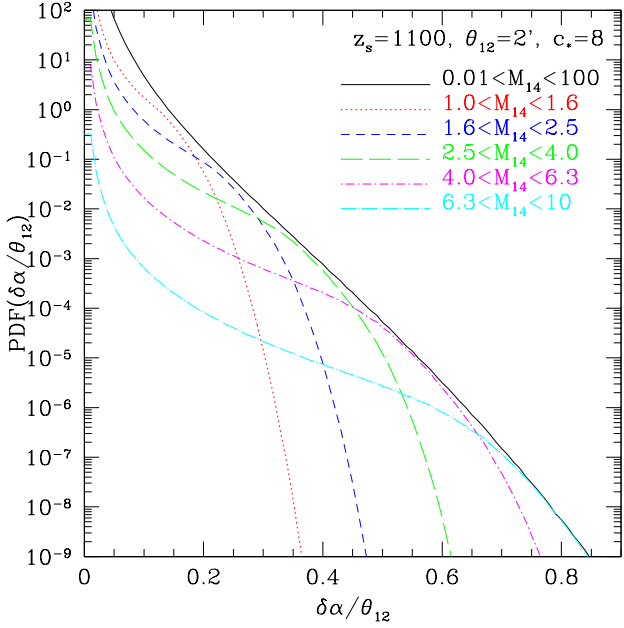}
\end{minipage}
\end{center}
\caption{Model prediction of the lensing excursion angle PDF for
$z_s=1100$, $c_*=8$ and $s=1$.
The black line shows the full PDF, while the colored lines show 
contribution from a limited range of the halo mass denoted in the panel.}
\label{fig:pdf_mrange}
\end{figure}

Under the assumptions stated in the last subsection, the excursion 
angle PDF is obtained by summing up contributions from single 
halos over a wide range of the halo mass and integrating over the 
redshift of halos as defined by eq.~(\ref{pdf}).
We plot the PDF computed from such a model in Figure \ref{fig:pdf_mrange}.
The black line shows the total PDF, while colored lines represent
contributions from narrow limited mass ranges.
This Figure clearly illustrates the origin of the exponential tail.
There are two key points; 
a large excursion angle can only be generated by
massive halos with mass typically larger than $10^{14}h^{-1}M_{\odot}$, 
and at such a halo mass range, the mass function decreases 
exponentially.
Accordingly, the number of more massive halos that can contribute to a
larger excursion angle decreases exponentially, and as a result, 
the exponential slope of the PDF arises.

%%%% Figure 10
\begin{figure}
\begin{center}
\begin{minipage}{8.2cm}
\epsfxsize=8.2cm 
\epsffile{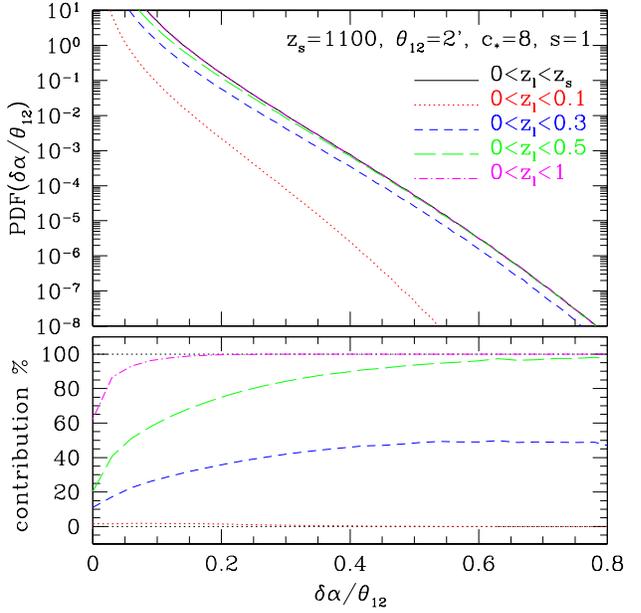}
\end{minipage}
\end{center}
\caption{Model prediction of the lensing excursion angle PDF for
$z_s=1100$, $c_*=8$ and $s=1$.
The black line shows the full PDF, while the colored lines show 
contribution from a limited range of the lens redshift denoted in 
the panel.}
\label{fig:pdf_zrange}
\end{figure}

In order to examine what redshift range of halos makes a major 
contribution to the exponential tail, 
we plot in Figure \ref{fig:pdf_zrange} the excursion angle PDFs 
computed for limited ranges of the lens redshifts (upper panel)
and their percentage of the contribution (lower panel).
It is seen in the lower panel that contributions from halos at redshifts 
below 1 account for almost full amplitude of the PDF tail.
This is explained by a rapid evolution of the halo mass function at the
high mass end.
In fact, the number density of massive halos with 
$M_{\rm halo}\ga 2\times 10^{14}h^{-1}M_\odot$ decreases by more than
one order of magnitude from $z=0$ to 1.

%%%% Figure 11
\begin{figure}
\begin{center}
\begin{minipage}{8.2cm}
\epsfxsize=8.2cm 
\epsffile{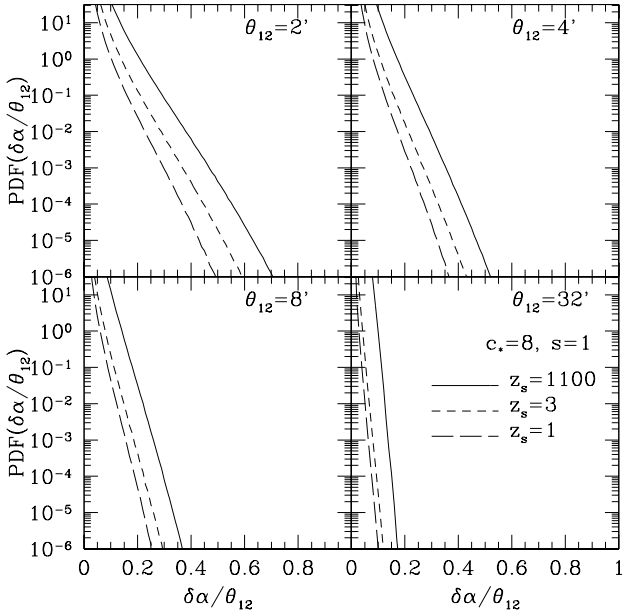}
\end{minipage}
\end{center}
\caption{The model predictions of the lensing excursion angle PDF
after the correction to the accumulative lensing effect being made
(using the approximate way of eq.~(\ref{econv})).
The halo parameters are $c_*=8$ and $s=1$.
The ray separations are denoted in each plot.
The source redshifts are, from inner to outer, $z_s$=1, 3 and 1100, 
respectively. }
\label{fig:pdf1side2x2z.gconv}
\end{figure}

The above results lead to the following explanation for the 
origin of the lensing
excursion-angle PDF which has two components, a Gaussian core and an
exponential tail:
The light rays emitted at high redshifts undergo many gravitational 
deflections by (large- or small-scale) structures on their way to us, 
which after many uncorrelated deflections produces the Gaussian core.
Some small part of the rays is strongly lensed by a massive halo with 
mass larger than $10^{14}h^{-1}M_\odot$ at a low redshift of $z<1$,
and the coherent deflections caused by the strong lensing produce the 
exponential tail. 
Therefore even if a ray pair encounters a strong coherent deflection 
by a single massive halo, its excursion angle is not solely determined 
by the strong halo lensing but the random scattering also contributes
to it to a smaller extent.
This effect on the PDF is taken into account by the convolution:
\begin{equation}
\label{conv}
E(x)=\int dy~G(y) E'(x-y),
\end{equation}
where $G(y)$ denotes the Gaussian distribution, and 
$E'(x)$ denotes the tail part produced by the halo lensing without
considering the contribution from the random scattering 
(which could be computed by the model described in this section).
Since, as shown in Figure \ref{fig:pdf_mrange}, the tail is well 
approximated by the exponential shape with a constant index over a wide 
range of $\delta \alpha$, it is reasonable to approximate 
$E'(x)\simeq p'\exp(-qx)$ and,
\begin{eqnarray}
\label{econv}
E(x)&\simeq&{p'\over{\sqrt{2 \pi} \sigma}} \int dy~
\exp\left(-{{y^2}\over{2\sigma^2}}\right) 
\exp(-q(x-y))\nonumber\\
&=& p \exp(-qx),
\end{eqnarray}
with the boosted amplitude, 
\begin{equation}
\label{pbias}
p=p' \exp\left({{(\sigma q)^2}\over{2}}\right).
\end{equation}
Thus under the above approximation, the slope of the exponential tail 
is unchanged but the amplitude is increased.
This combined with the fact that almost all contributions to the PDF 
tail come from coherent scattering by massive halos at $z<1$ 
(Figure \ref{fig:pdf_zrange}) accounts 
for the trend observed in the excursion angle
PDFs obtained from the ray-tracing simulation that its slope does not 
depend strongly on the source redshift but
its amplitude increases with the source redshift.
Actually, a very similar trend is observed in the model PDFs plotted in 
Figure \ref{fig:pdf1side2x2z.gconv} which shows the corrected PDF tails for 
$z_s=1$, 3 and 1100 and for four ray separations.
Here, in order to compute the boost factor of eq.~(\ref{pbias}), we compute 
$p'$ and $q$ by fitting the model PDFs 
at two points, $\pdf(x1)=1.0\times10^{-2}$ and $\pdf(x2)=1.0\times10^{-3}$
to the exponential function.
Note that after this correction the amplitude of the tails can be 
increased more than one order of magnitude because of the steep slope 
of the exponential tail (thus for a large $q$).

%%%% Figure 12
\begin{figure}
\begin{center}
\begin{minipage}{8.2cm}
\epsfxsize=8.2cm 
\epsffile{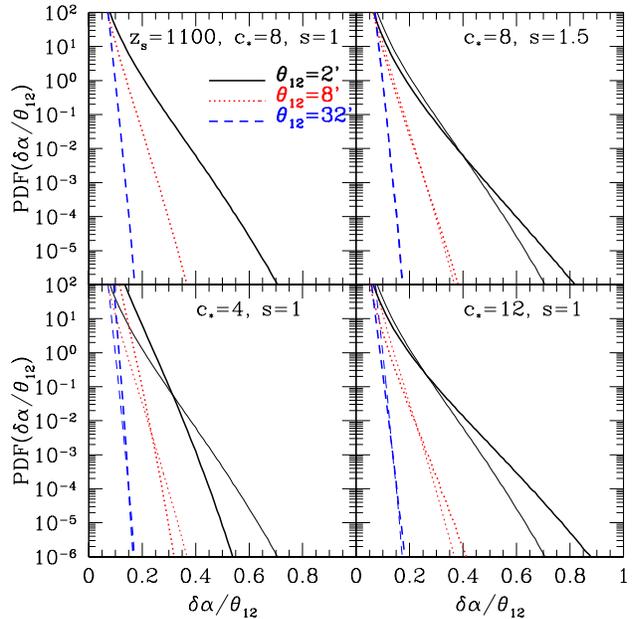}
\end{minipage}
\end{center}
\caption{The model predictions of the lensing excursion angle PDFs.
We have applied the corrections given by equations (\ref{econv}) and 
(\ref{pbias}).
Ray separations are $\theta=2$, 8 and 32 arcmin (solid, dotted and dashed,
respectively).
The source redshift is taken by $z_s=1100$.
Halo model parameters are denoted in each panel.
The thin lines in the top right and two bottom panels show,
for comparison, the predictions of the fiducial model plotted in the 
top left panel.}
\label{fig:pdfcompari.gconv}
\end{figure}

Four panels of Figure \ref{fig:pdfcompari.gconv} show the corrected model
PDFs for various values of $c_*$ and $s$, which help to understand the 
dependences of the shape of the PDF tail on the halo parameters.
Clearly, the broader tail appears for models with a larger $c_*$ or a larger 
$s$, because such models generate a larger maximum deflection angle. 
It is important to notice that a small change of $c_*$ or $s$ causes 
a very large, nonlinear change in the shape and amplitude of the PDF.
Therefore, choosing mean values of $c_*$ and $s$ does not provide
a mean PDF but gives a lower amplitude.
It should be noted that the halo model parameters indeed have a large 
scatter (Jing 2000; Jing \& Suto 2000; see also Figure 9 of 
Hamana et al.~(2004) which clearly shows that the compactness of 
halo mass distributions has a large scatter).

\begin{figure}
\begin{center}
\begin{minipage}{8.2cm}
\epsfxsize=8.2cm 
\epsffile{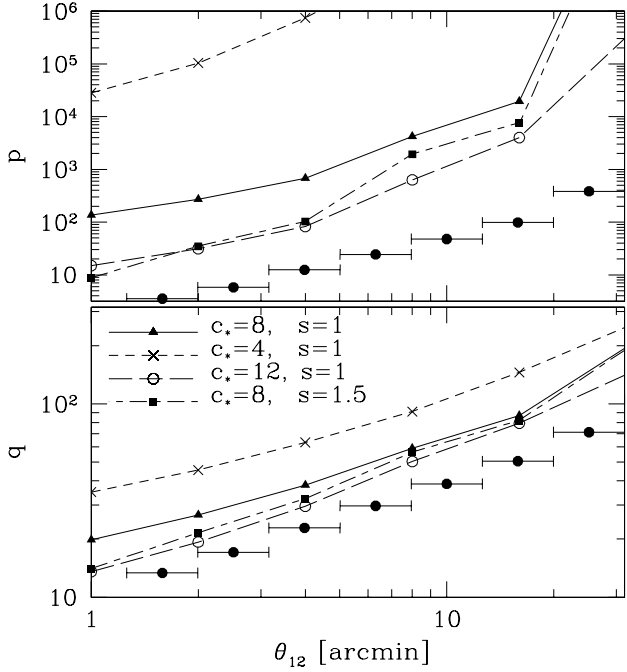}
\end{minipage}
\end{center}
\caption{Parameters in the exponential distribution (eq. \ref{econv}) 
computed from the fit to the model predictions of excursion angle PDFs
for $z_s=1100$.
Model parameters are denoted in the plot.
Dotted lines show the results from ray-tracing simulation.}
\label{fig:pqfit_emodel}
\end{figure}

Finally, we compare in Figure \ref{fig:pqfit_emodel} the parameters in the
exponential function eq.~(\ref{econv}) measured from the model PDFs 
(symbols with lines), with the results from the ray-tracing experiments.
We evaluate $p$ and $q$ by fitting the corrected model PDF (i.e.,
after the correction by eq.~(\ref{econv}) being made).
As shown in Figure \ref{fig:pqfit_emodel}, the parameter $q$ correlates 
with the slope parameter $q$ as expected.
The measured exponential slope parameter $q$ plotted in the lower panel 
are larger than the results from the numerical experiments, though the 
slope with the separation angle is very similar.
The discrepancy is smaller for models with a larger $c_*$ or a larger $s$.
Therefore the model prediction may be improved if one takes into account the 
scatter in the halo parameters $c_*$ or a larger $s$.

%%%% Figure 13

Quantitatively, none of the four models plotted in Figures
\ref{fig:pdfcompari.gconv} and \ref{fig:pqfit_emodel} are in very good
agreement with the simulation results.  The discrepancy is partly due to the
scatter in the halo model parameters as has been discussed above.  Also, a
deviation in the halo mass distribution from spherical symmetry could partly
account for it.  Actually, the mass distribution of most of the halos
significantly deviates from spherical symmetry (Hamana et al.~2004).  In
addition, the spatial correlation of halos may have an influence on the
excursion angle PDF, because massive halos are strongly clustered.  A close
look at the sky distributions of ray pairs having a large excursions angle and
of massive halos shown in Figure \ref{fig:rays_halos} reveals that those
deviations from our simple model should indeed play a role; namely, it is seen
in the Figure that a small part of most massive halos does not produce a large
lensing excursion event, and that a small part of large excursion angle ray
pairs does not intersect a very massive halo.

We may conclude, from what has been seen above, that our simple model succeeds
in getting the essential mechanism of generating the exponential tail and in
explaining the origin of the major characteristics of the tail.  The model
predictions are in reasonable agreement with the simulation results.
Further modifications of the model taking into account details of halo
properties, such as scatter in the halo model parameters, deviations of the
halo mass distribution from spherical symmetry and clustering of halos, 
are needed to improve the accuracy of the model prediction..

%%%%%%%%%%%%%%% sec.4 Summary and discussions
\section{Summary and discussions}

We have investigated the statistical distribution of lensing excursion
angles, paying a special attention to the physical processes that are 
responsible for generating two components of the PDFs: the Gaussian core
and the exponential tail.
We have used the numerical gravitational lensing experiments in a 
CDM cosmology to quantify these two components.

The origin of the Gaussian core is explained by the random lensing 
deflections by either linear or nonlinear structures.
The variances of the Gaussian core measured from the results of the 
numerical experiments are found to be in a good agreement
with the prediction by the power spectrum approach (Seljak 1994; 1996).

The presence of the exponential tail was first found by Hamana \& Mellier
(2001) but its origin remains unrevealed.
The tail is characterized by two parameters: the slope and 
amplitude. 
We have found from the numerical experiments that
the slope changes little with the source redshift 
while the amplitude becomes greater as the source redshift increases,
at least within the redshift range we consider ($1<z_s<1100$).
Since the random lensing deflections result in the Gaussian core, 
the exponential tail is most likely to result from coherent deflections.
In addition, in order to generate a large excursion angle, massive 
virialized objects should be responsible for the exponential tail.
Therefore, we supposed that the exponential tail originates from 
coherent lensing scatters by single massive halos.

We have developed a simple empirical model for the exponential tail of the 
lensing excursion angles PDF.
We used the analytic models of the dark matter halos, namely the 
modified Press-Schechter (Press \& Schechter 1974) mass function 
(Sheth \& Tormen 1999) and 
the Universal density profile first proposed by Navarro et al.~(1996).
Although we only consider a coherent lensing scatter by a single massive 
halo and did not take into account the scatters in the halo parameters
(the concentration parameter $c_\ast$ and the inner slope $s$), our model 
reasonably reproduces the exponential tails computed 
from the numerical experiments.
It is found that the massive halos with $M>10^{14}h^{-1}M_\odot$ are 
responsible to the tail and that the exponential shape arises as a 
consequence of the exponential cutoff of the halo mass function at 
such mass range.
Almost all contributions to the tail come from the halos at redshifts
below 1.
Therefore, the slope of the tail is formed by the halos at $z<1$.
On the other hand, the amplitude of the tail is determined by the 
convolution of two contributions, the coherent scatter and the random 
deflections.
Since the contribution from the random deflections becomes greater
as the source redshift increases, the amplitude of the tail becomes 
greater for a higher source redshift.
These explain the redshift-independent slope and the redshift-dependent
amplitude found from the numerical experiments.

Does the exponential tail have an influence on the angular power spectrum
of the temperature map of the cosmic microwave background ($C_\ell$) ?
As far as angular scales larger than 1 arcmin are concerned, 
the answer is no.
When one computes the lensed $C_\ell$ one can safely use the approximate
convolution equation given by Seljak (1996), because the key assumption 
in the approximation made for deriving the convolution 
equation is not the Gaussianity of the excursion angle PDF, but that its 
variance is small (cf.~\S 9 of Bartelmann \& Schneider 2001).
Since the smallness of the variance is also the case for lower redshifts,
the same convolution technique can be applied to other angular correlation 
functions such as that of galaxies and QSOs.

%%%% Figure 14
\begin{figure}
\begin{center}
\begin{minipage}{8.2cm}
\epsfxsize=8.2cm 
\epsffile{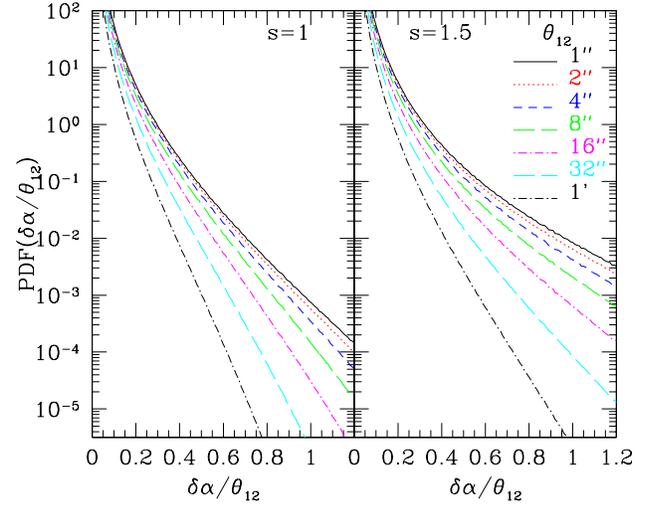}
\end{minipage}
\end{center}
\caption{The empirical model predictions of the tail part of the 
excursion angle PDF for sub-arcmin ray separations.
The source redshift is $z_s=1100$, the concentration parameter is
$c_\ast=8$ and the inner slope is $s=1$ (left) and $s=1.5$ (right).}
\label{fig:pdfones}
\end{figure}

Before closing this paper, we present predictions for tails of the 
excursion angle PDF for sub-arcmin ray separations.
We plot in Figure \ref{fig:pdfones} our empirical model predictions.
It is found that for ray separations $\theta_{12}>10$ arcsec, the 
amplitude of the tail keeps increasing with the slope becoming flatter 
in a similar rate of larger separations.
However below that separation, the growth rate becomes smaller gradually.
The standard dispersions of these distributions are smaller than unity 
but can be of order $O(0.1)$ which is comparable to that of the Gaussian core, 
though our simple model adopting average halo parameters tends to predict
greater amplitude than the results from the numerical experiments
(see \S 3.4).
Therefore, it is possible that on arcsecond scales coherent lensing 
deflections have non-negligible influence on the angular clustering 
of objects in the distant universe.
Note that even if taking the exponential tail into account, the standard 
dispersion of the excursion angles is less than unity, thus the approximate 
convolution equation can be still valid but the contribution from the tail to
the dispersion should be included.
We notice that it is however not clear whether the assumptions in our model 
are still valid on such small ray separations.
The statistical distribution of lensing excursion angles for arcsecond
separation ray pairs should be investigated in a future work with
a gravitational numerical experiment having a higher resolution.

%%%%%%%%%%%%%%% Acknowledgments
\section*{Acknowledgments}

The $N$-body simulations used in this work were carried out by the 
Virgo Consortium at the computer center at Max-Planck-Institut, Garching
(http://www.mpa-garching.mpg.de/NumCos). 
Numerical computations presented in this paper were partly
carried out at ADAC (the Astronomical Data Analysis Center) of the
National Astronomical Observatory Japan,  at the 
Yukawa Institute Computer Facility and at the computing center of the
Max-Planck Society in Garching.

%%%%%%%%%%%%%%% References

\label{lastpage}
\end{document}